\documentclass{svjour3}
\RequirePackage{amsmath,amssymb,graphicx}
\RequirePackage{graphics}

\journalname{Eur. Phys. J. C}

\usepackage{caption}
\usepackage{dcolumn}
\usepackage{bm}
\usepackage{color}
\usepackage{graphicx}

\newcommand{\La}{$\Lambda$}

\newcommand{\pip}{$\pi^{+}$}
\newcommand{\pim}{$\pi^{-}$}
\newcommand{\kap}{K$^{+}$}
\newcommand{\kam}{K$^{-}$}

\newcommand{\sNN}{$\sqrt{s_{\rm NN}}$}
\newcommand{\be}{\begin{equation}}
	\newcommand{\ee}{\end{equation}}

\def\muB{$\mu_B$}

\begin{document}
	\title{Thermal Model Description of Collisions of Small Nuclei}
	\author{H.~Oeschler
\and
J.~Cleymans
\and
B.~Hippolyte
\and
K.~Redlich
\and
N.~Sharma
}
\institute{H.~Oeschler\at Physikalisches Institut, Ruprecht-Karls-Universit\"at Heidelberg, D-69120 Heidelberg, Germany
	  \and J.~Cleymans
	 \at UCT-CERN Research Centre and Department  of  Physics,\\ University of Cape Town, Rondebosch 7701, South Africa
	\and B.~Hippolyte
	 \at Institut Pluridisciplinaire Hubert Curien (IPHC), Universit\'e de Strasbourg, CNRS-IN2P3, F-67037 Strasbourg, France
	\and K.~Redlich
	\at Institute of Theoretical Physics, University of Wroc\l aw, Pl-50204 Wroc\l aw, Poland
	\at ExtreMe Matter Institute EMMI, GSI, D-64291 Darmstadt, Germany
	\and N.~Sharma
	\at {Department of Physics, Panjab University, Chandigarh 160014, India}
	}
	\date{\today}

%
%
\maketitle

\begin{center}
{\sl Dedicated to the memory of Helmut Oeschler}
\end{center}
\begin{abstract}
The  dependence of particle production on the size of the colliding nuclei is
analysed in terms of the  thermal model using the  canonical ensemble.
The concept of strangeness correlation in clusters of sub-volume $V_c$ is used
to account for the suppression of strangeness.
A systematic analysis is presented of the predictions of the thermal model
for particle production in collisions of small nuclei.
The pattern of the maxima of strange-particles-to-pion ratios as
a function of beam energy is quite special, as they do not occur  at the same beam energy
and are sensitive to system size. In particular,
the $\Lambda/\pi^+$ ratio shows a clear maximum even for small systems
while the maximum in the K$^+/\pi^+$ ratio 
 is less pronounced
in small systems.
\end{abstract}

\section{Introduction}
A substantial experimental effort is presently under way to study not only heavy- but also light-ion
collisions. This is being motivated by the puzzling results, obtained in Pb-Pb and Au-Au collisions,  for
the non-monotonic  behavior of the \kap/\pip ratio,  and other particle ratios which
have been  conjectured as being due to  a phase change  in nuclear matter~\cite{Gazdzicki:1998vd}.

A consistent description of particle production in heavy-ion collisions, up to LHC energies,
has emerged during the past two decades using a thermal-statistical model (referred to simply as thermal model in the remainder of this paper).
It is based on the creation and subsequent decay of hadronic resonances produced in chemical
equilibrium at a unique temperature and baryo-chemical potential.  According to this picture the bulk
of hadronic resonances made up of the light flavor (u,d and s) quarks are  produced in chemical equilibrium.

Indeed, some particle ratios exhibit very interesting features as a function of  beam energy  which deserve
attention: (i) a maximum in the \kap/\pip ratio, (ii) a maximum in the $\Lambda/\pi$ ratio, (iii) no maximum in the \kam/\pim ratio.
These three features  occur  at a centre-of-mass energy of around 10  GeV~\cite{Andronic:2005yp,Cleymans:2005xv,Cleymans:2004hj}.
The 
 maxima happen in an energy regime where the largest net
baryon density occurs~\cite{Randrup:2006nr,randrup2016} and a transition from a baryon-dominated freeze out to a
meson dominated one takes place~\cite{Cleymans:2004hj}. An alternative interpretation
is that these maxima reflect a phase change to a deconfined state of matter ~\cite{Gazdzicki:1998vd}.

The maxima mark a distinction between heavy-ion collisions and p-p collisions as they have not been observed in the latter.
This shows a clear difference between the two systems which is worthy of further investigation.

It is the purpose of the present paper to investigate the transition from a small system like a p-p collision to a large system
like a Pb-Pb or Au-Au collision and to follow  explicitly the genesis of the maxima in certain particle ratios.
This is relevant for the interpretation of the data coming out of the BES~\cite{Mustafa:2015yeg} and NA61~\cite{Abgrall:2014xwa} experiments. These
experiments are  spearheading this effort at the moment,  in the
near future additional results will be obtained at the NICA collider
and at the FAIR  facility.

\section{The model}

A relativistic heavy-ion collision goes through several stages.
At one of the later stages, the system is dominated by
hadronic resonances.
The identifying feature of the thermal model  is that all the resonances as listed by the Particle Data Group~\cite{Agashe:2014kda} are
assumed to be in thermal and chemical equilibrium.
This  assumption drastically reduces the number of free parameters and thus this stage is determined by just a few
thermodynamic variables namely, the chemical freeze-out temperature $T$, the various chemical potentials $\mu$ determined by
the conserved quantum numbers and by the volume $V$ of the system.
It has been shown that this description is also the correct
one~\cite{Cleymans:1997eq,Akkelin:2001wv,Broniowski:2001we} for a scaling expansion as first discussed by
Bjorken~\cite{Bjorken:1982qr}.

In general, if the number of particles carrying quantum numbers related to a conservation law is small, then
the grand-canonical description no longer holds. In such a case, conservation of charges has to be implemented
exactly by using the canonical ensemble~\cite{BraunMunzinger:2001as,BraunMunzinger:2003zd,Cleymans:1998yb,Hamieh:2000tk}.
We start by presenting a brief reminder of the general concepts of the thermal model.
\subsection{Grand Canonical Ensemble}
In the  grand-canonical  ensemble, the volume $V$, temperature $T$  and the  chemical
potentials $\vec \mu$
determine the partition function $Z(T,V,\vec\mu)$. In the hadronic fireball of non-interacting
hadrons, $\ln Z$ is the sum of the contributions of all $i$-particle species
given by
\begin{equation}
\frac{1}{V} \, {\rm ln}Z(T, V, \vec{\mu}) = \sum_i {}Z_i^1(T, {\vec\mu}), \label{equ1}
\end{equation}
where  $\vec{\mu}=(\mu_B,\mu_S,\mu_Q)$ are
the chemical potentials related to the conservation of baryon number, strangeness and electric charge, respectively.

The partition function  contains all information needed to obtain the number density $n_i$ of
particle species $i$. Introducing the particle's specific chemical potential $\mu_i$, one gets
\begin{equation}
n_i^{}(T, \vec{\mu}) = \frac{1}{V}\left. \frac{\partial(T \ln Z)}{\partial\mu_i}\right|_{\mu_i=0}. \label{equ3}
\end{equation}
Any resonance that decays into species $i$  contributes to the yields eventually measured. Therefore,
the contributions from all heavier hadrons $j$ that decay to hadron $i$ with the branching fraction
$\Gamma_{j \rightarrow i}$ are given by:
\begin{equation}
n_i^{\rm decay} = \sum_j  \Gamma_{j \rightarrow i} ~ n_j. \label{equ4}
\end{equation}
Consequently, the final yield $N_i$ of particle species $i$ is the sum of the thermally
produced particles and the decay products of resonances,
\begin{equation}
N_i = (n_i^{} + n_i^{\rm decay}) ~ V. \label{equ5}
\end{equation}
From Eqs.~(\ref{equ3}-\ref{equ5}) it is clear that in  the grand-canonical ensemble the particle yields are determined by the
volume of the fireball, its temperature and the chemical potentials.

\subsection{Canonical ensemble}
If the number of particles  is small, then
the grand-canonical description no longer holds. In such a case conservation laws have to be implemented
exactly.
Here, we refer only  to strangeness conservation and
consider charge and baryon number conservation to be fulfilled on the average in the grand canonical
ensemble because the number of charged particles and baryons is much larger than that of strange
particles~\cite{BraunMunzinger:2001as}.
The density of strange particle $i$ carrying strangeness $s$ can be obtained from
(see~\cite{BraunMunzinger:2001as} for further details),
\begin{eqnarray}
n_{i}^C&=&{\frac{Z^1_{i}}{Z_{S=0}^C}} \sum_{k=-\infty}^{\infty}\sum_{p=-\infty}^{\infty} a_{3}^{p} a_{2}^{k}
 a_{1}^{{-2k-3p- s}} 
I_k(x_2) I_p(x_3) I_{-2k-3p- s}(x_1),   \label{equ6}
\end{eqnarray}
where $Z^C_{S=0}$ is the canonical partition function
\begin{eqnarray}
Z^C_{S=0}&=&e^{S_0} \sum_{k=-\infty}^{\infty}\sum_{p=-\infty}^{\infty} a_{3}^{p}
a_{2}^{k} a_{1}^{{-2k-3p}} 
I_k(x_2) I_p(x_3) I_{-2k-3p}(x_1),
\label{eq7}
\end{eqnarray}
and $Z^1_i$ is the one-particle partition function  (Eq.~\ref{eq7}) calculated for $\mu_S=0$ in the Boltzmann
approximation. The arguments of the Bessel functions $I_s(x)$ and the parameters $a_i$ are introduced as,
\begin{eqnarray} a_s= \sqrt{{S_s}/{S_{\mathrm{-s}}}}~~,~~ x_s = 2V\sqrt{S_sS_{\mathrm{-s}}} \label{eq8a}, \end{eqnarray}
where $S_s$ is the sum  of all $Z^1_k$  for particle species $k$ carrying strangeness
$s$.

In the limit where $x_n<1$ (for $n=1$, 2 and 3)   the density of strange particles carrying
strangeness $s$ is well approximated by~\cite{BraunMunzinger:2001as}
\begin{equation}
n_i^{C} \simeq n_i \frac{I_{s}(x_1)}{I_0(x_1)}. \label{equ7}
\end{equation}
From these  equations it is clear that in the canonical ensemble the strange particle density depends explicitly on the
volume  through the arguments of the Bessel functions.
This volume might be different from the overall volume $V$ and is denoted as $V_c$~\cite{Cleymans:1998yb,Hamieh:2000tk,satz}.


%
%


In this case there are two volume parameters: the overall
volume of the system $V$, which determines the particle yields at fixed density and the
strangeness correlation volume $V_c$, which describes the range of strangeness conservation. If this volume is small,
it 
reduces the densities of strange particles.
%


\section{Origin of the maxima}

It has been observed that the baryon chemical potential decreases  with increasing beam energy
while  the temperature increases  quickly and reaches  a plateau.
Following the rapid rise of the temperature at low beam energies, the $\Lambda/\pi^+$
and  \kap/\pip also increase rapidly.
This halts when the temperature reaches its limiting value. Simultaneously the baryon chemical potential keeps on decreasing.
Consequently, the  $\Lambda/\pi$ and \kap/\pip ratios follow this decrease due to
 strangeness conservation as \kap~is produced in associated production together with a \La.
The two effects combined lead to maxima in both cases.
For very high energies, the baryo-chemical potential no longer plays a role ($\mu_B\approx 0 $) and, since the temperature remains constant,
these ratios no longer  vary~\cite{Cleymans:2004hj}.

In order to analyse the strangeness content in a heavy ion collision we make use of the
Wroblewski factor~\cite{Wroblewski:1985sz} which is defined as
$$
\lambda_s = \frac{2\left<s\bar{s}\right>}{\left<u\bar{u}\right> + \left<d\bar{d}\right>}  .
$$
This factor is determined from the quark content of the hadronic resonances, namely,
by the number of  newly created  strange -- anti-strange ($\left<s\bar{s}\right>$) and the
non-strange $\left<u\bar{u}\right>$ and $\left<d\bar{d}\right>$ quarks
{\it before} their strong decays.

\begin{figure}
\centering
\includegraphics[clip,trim=0 1cm 0 11cm,width=0.65\textwidth, height=5.5cm]{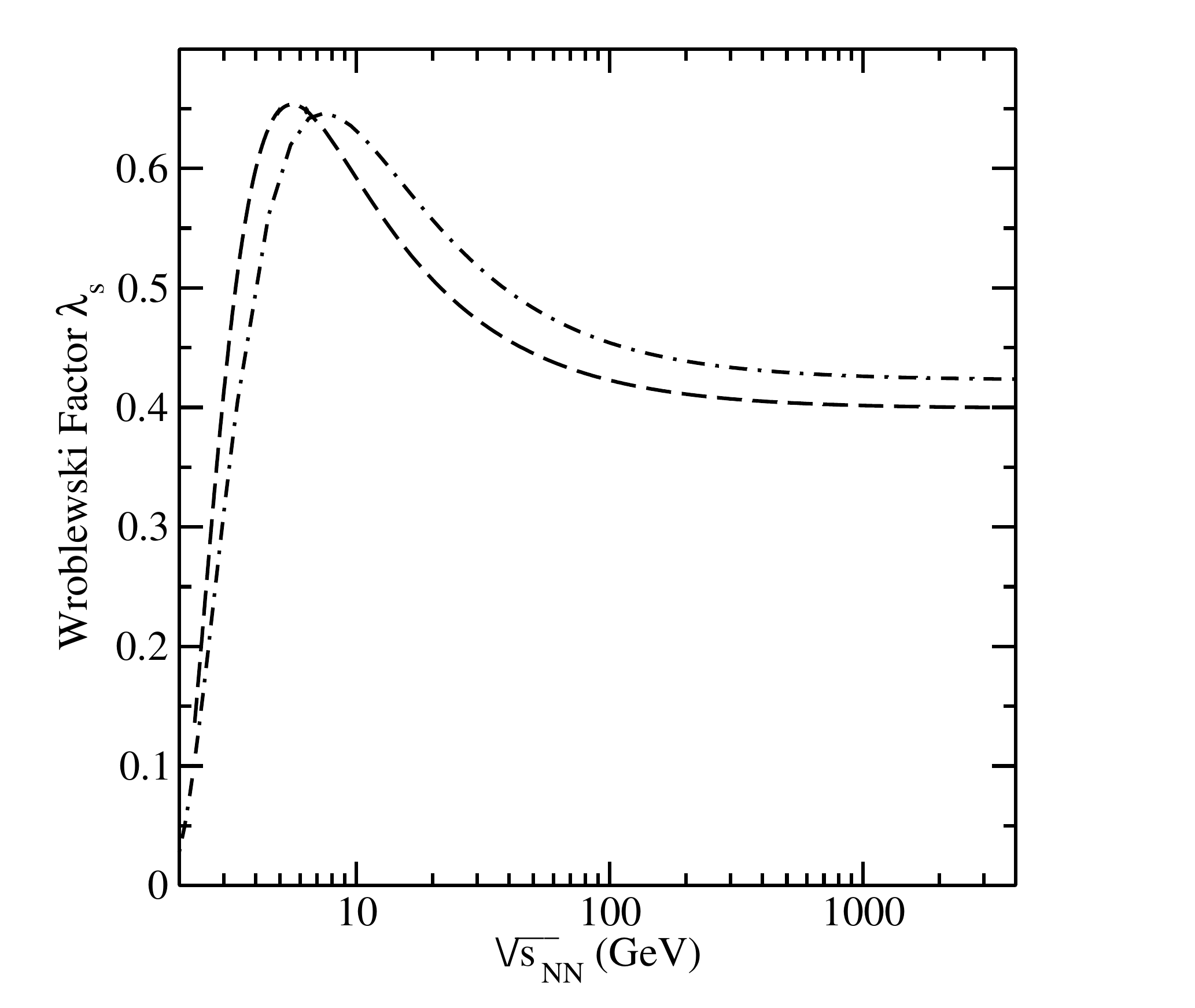} 	
  \caption{The Wroblewski factor $\lambda_s$ as a function of beam energy calculated along the
		chemical freeze-out curve.
The dashed-dotted line is calculated along  the freeze-out curve obtained in~\cite{Cleymans:2005xv} while the dashed line
uses the parametrization given in~\cite{Vovchenko:2015idt}.}
	\label{fig:wrob}
\end{figure}

Its limiting values are obvious:
$\lambda_s = 1$, if all quark pairs are equally abundantly produced, i.e. flavor SU(3) symmetry and
$\lambda_s = 0$, if no strange quark pairs are present in the final state.
This factor has been calculated in the thermal model using the THERMUS~\cite{Wheaton:2004qb} code by examining the quark content of hadronic resonances.
Due to its definition, contributions from heavy flavors like charm are explicitly excluded.
The relative strangeness production
in heavy-ion collisions along the chemical freeze-out line shows a
maximum.  This is illustrated by the Wroblewski factor in Fig.~\ref{fig:wrob}.
The functional dependence of $T$ and \muB{} on \sNN{} as
in \cite{Cleymans:2005xv} was used.
For comparison we also show  the freeze-out parameterization recently introduced in~\cite{Vovchenko:2015idt},  
which has a lower limiting temperature than the one given in~\cite{Cleymans:2005xv}.

In an earlier publication, we have already discussed  how  the maximum in $\lambda_s$,  seen in Fig.~\ref{fig:wrob},   
 occurs
~\cite{BraunMunzinger:2001as}.
It turns out that the shape of the Wroblewski factor tracks the energy dependence of  the K$^+/\pi^+$ ratio.

\begin{figure}[h]
\includegraphics[width = 0.55\textwidth, height=5.2cm]{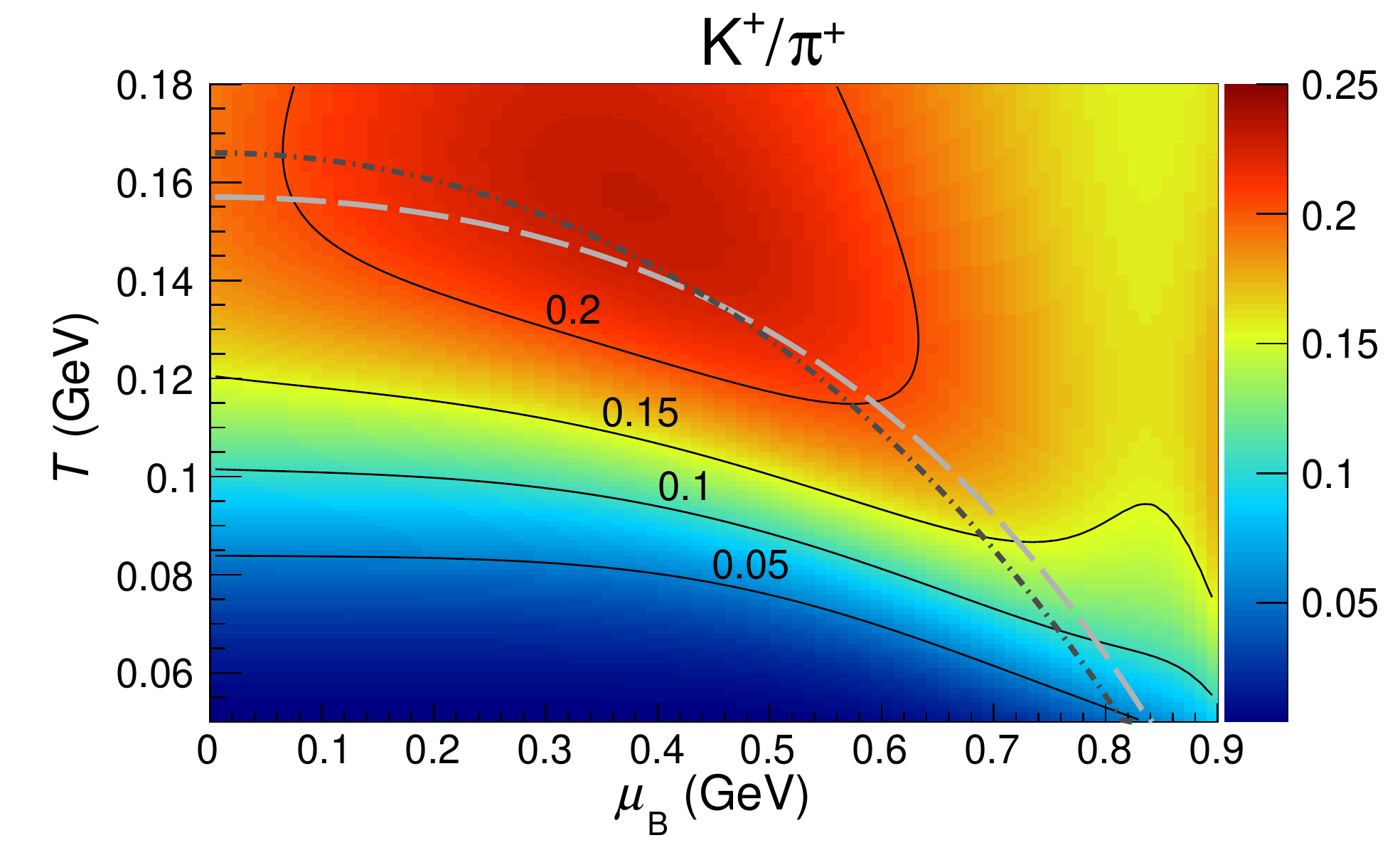}
\hfill\includegraphics[width = 0.55\textwidth, height=5.2cm]{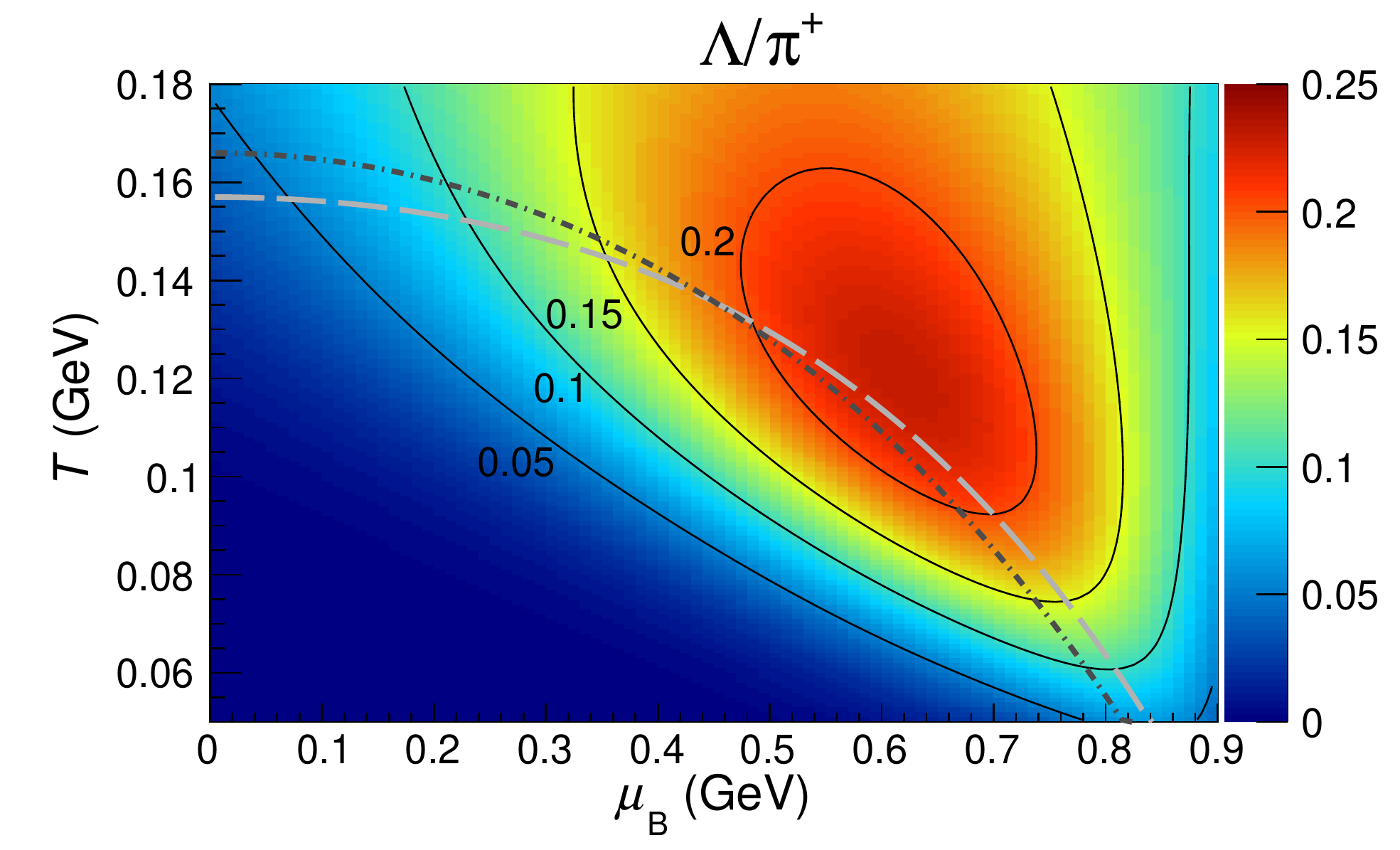}
\caption{Values of the \kap/\pip (left-hand pane) and the $\Lambda/\pi^+$ (right-hand pane) ratios in the $T-\mu_B$ plane.
Lines of constant values are indicated.
The dashed-dotted line is the freeze-out curve obtained in~\cite{Cleymans:2005xv} while the dashed line
uses the parameterization given in~\cite{Vovchenko:2015idt}.
Note that the maxima do not occur in the same position.
 }
\label{fig:constant}
\end{figure}

To show this in more detail we present as an example in Fig.~\ref{fig:constant}  contour lines where the \kap/\pip  and the
$\Lambda/\pi^+$ ratios remain constant in the $T-\mu_B$ plane.
It should be noted that the  maxima of these ratios do not  occur in the
same position, which remains to be confirmed experimentally. It is also worth noting that in these cases the maxima
are not on,  but slightly above  the freeze-out curve.

\vspace{1cm}
\vspace*{1cm}

\section{Particle ratios for small systems}

To consider the case of the collisions of smaller nuclei we take into account
the strangeness suppression using  the concept of strangeness correlation
in clusters of a sub-volume $V_c\leq V$~\cite{Cleymans:1998yb,Hamieh:2000tk,Kraus:2007hf}.

A particle with
strangeness quantum number $s$ can appear anywhere in the volume $V$ but it has to
be accompanied by another particles carrying strangeness $-s$
to conserve strangeness in the correlation volume $V_c$ .
Assuming spherical geometry, the volume $V_c$ is parameterized by the radius $R_c$ which is a free
parameter that defines the range of local strangeness equilibrium.

In the following we show the trends of various particle ratios as a function of \sNN.
The  dependence of $T$ and \muB{}  on the beam energy is taken from  heavy-ion collisions~\cite{Cleymans:2005xv}.
For p-p collisions  slightly different parameters would be  more suited~\cite{Cleymans:2011pe}.
Therefore, the calculations illustrate the general trend, as we have ignored the variations of the parameters with system size.

We focus on the system-size dependence of the thermal parameters with particular emphasis on
the change in the strangeness correlation radius $R_c$. The radius parameters of the volume V, $R$  = 10 fm (which is
the value for central Pb-Pb collisions) and $\gamma_S$ = 1  are kept fixed. The freeze-out values
of $T$ and $\mu_B$ will vary with the system size~\cite{Kraus:2007hf}, however this has not been taken
into account in the present work
which therefore gives only a qualitative description of the effect.

The smaller system size is described by decreasing the value of the correlation radius $R_c$. This ensures that strangeness
conservation is exact in $R_c$, and hence, strangeness production is more suppressed with
decreasing $R_c$.

\begin{figure}[h]
\centering
\includegraphics[width = 14cm]{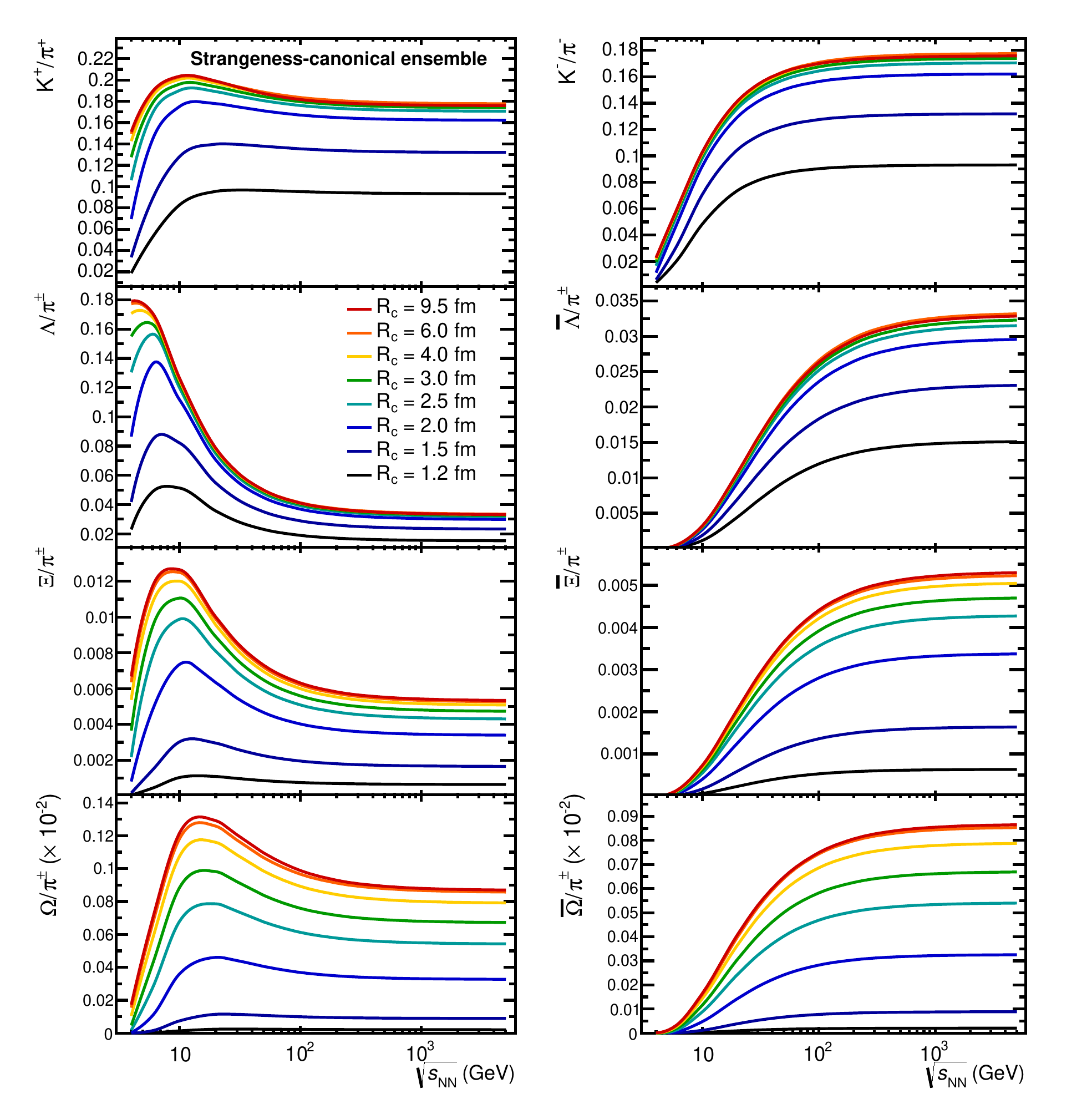} 		

  \caption{Behavior of particle ratios  as a function of the invariant beam energy for various
strangeness correlation radii $R_c$, calculated using the
thermal model  formulated in  a
canonical ensemble \cite{Wheaton:2004qb}. The  \kap/\pip,  $\Lambda/\pi^\pm$, $\Xi/\pi^\pm$ and
$\Omega/\pi\pm$ ratios are shown in the left-hand panel
while the corresponding antiparticles are shown in the right-hand panel.
Note that the $\Lambda/\pi^\pm$ ratio is the only ratio where the maximum does not disappear as the system size is
reduced.}
  \label{ratio-p}
\end{figure}

In Fig.~\ref{ratio-p}  we show the energy and system size dependence of different particle ratios calculated along
the chemical freeze-out line. A maximum is seen in the \kap/$\pi^+$ ratio which gradually disappears when the correlation radius
 decreases.
A different  effect is seen in   $\Lambda/\pi^\pm$ ratio. Here, the gradual decrease
of the maximum is also seen  but, contrary to the K$^+/\pi^+$ ratio, it  does not  disappear and is still present
even for  small radii.
This behavior can be tested  experimentally and, if confirmed,  will
give support the hadronic scenario presented here. \\[0.4cm]

In the left-hand panel of Fig.~\ref{ratio-p}  it is also  seen that
for different particle ratios the maxima gradually become less pronounced
 as the size of the system decreases.
Also, the maximum shifts, for smaller systems, towards higher \sNN. For pp collisions which
correspond to a $R_c$ of about 1.5 fm~\cite{Kraus:2007hf}, they will hardly be observed, except for $\Lambda/\pi$ ratio.
As noted before the  maxima happen at  different beam energies.

The corresponding ratios for antiparticles are shown in the right-hand panel of Fig.~\ref{ratio-p}.
As is to be expected in the thermal model,
no maxima are present
because the baryon chemical potential $\mu_B$ enters with the opposite sign
and the ratios increase smoothly
with increasing beam energies until they reach a constant value corresponding to the limiting
hadronic temperature.

The ratios involving multi-strange baryons are also shown in~Fig.~\ref{ratio-p}. It is to be noted here that the
maxima occur at a higher beam energy than for the K$^+/\pi^+$ and the $\Lambda/\pi$ ratios.  {\it The maxima are caused by an 
interplay of different mass thresholds,  a decreasing $\mu_B$ and the saturation of $T$.}
The maxima gradually disappear as the size of the system is reduced.
These are the main results of the present paper.

It must be emphasized that the results presented here are of a qualitative nature. In particular there could be changes
due to  variations with the system size of the temperature and the baryon chemical potential. In addition the strangeness equilibration volume $V_c$ could be energy dependent and system-size dependent.

\section{Conclusions}

The thermal model describes the presence of maxima in the \kap/\pip and the $\Lambda/\pi^\pm$ ratios
at a beam energy of \sNN $~\approx$ 10  GeV.
In this paper we have
described what could possibly happen
with different strange particles and pion yields in collisions of smaller systems due to constraints imposed by
exact strangeness conservation. To this end,  use was made of a correlation volume to account
for the strangeness suppression effect.
We have shown that, in general, the characteristic feature of such ratios is a non-monotonic excitation
function with well identified maxima for the ratios
involving strange particles. The ratios with the anti-strange particle yields exhibit a monotonic increase and saturation with energy.
A decrease in the maxima was noted and for certain ratios of particle yields
the maxima completely disappear but not for all.
In particular, the $\Lambda/\pi^+$ ratio still shows a clear maximum even for  small systems.
The pattern of these maxima is also quite special, they are not always at the same beam energy.

If all ratios are following the trend  given here, it is a strong argument that the
properties of the  strange particle excitation functions, and their system size dependence,
 are
governed by the hadronic phase of the collisions constrained by an exact strangeness
conservation implemented in a canonical ensemble.

\begin{acknowledgements}
The work of K.R. was supported  by  the Polish Science
 Center (NCN), under Maestro grant DEC-2013/10/A/ST2/00106. 
N.S. acknowledges the support of DST-SERB Ramanujan Fellowship (D.O. No. SB/S2/RJN- 084/2015).
\end{acknowledgements}


\end{document}